# Mid-infrared frequency combs

Albert Schliesser[1,2], Nathalie Picqué[1,3,4,*], Theodor W. Hänsch[1,3]

[1] Max-Planck Institut für Quantenoptik, Hans-Kopfermann Strasse 1, D-85748 Garching, Germany
[2] École Polytechnique Fédérale de Lausanne (EPFL), CH-1015 Lausanne, Switzerland
[3] Ludwig-Maximilians-Universität München, Fakultät für Physik, Schellingstrasse 4/III, 80799 München, Germany
[4] Institut des Sciences Moléculaires d'Orsay, CNRS, Bâtiment 350, Université Paris-Sud, 91405 Orsay, France

* author to whom correspondence should be addressed, nathalie.picque@mpq.mpg.de

**Abstract**

Laser frequency combs are coherent light sources that emit a broad spectrum consisting of discrete, evenly spaced narrow lines, each having an absolute frequency measurable within the accuracy of an atomic clock. Their development, a decade ago, in the near-infrared and visible domains has revolutionized frequency metrology with numerous windfalls into other fields such as astronomy or attosecond science. Extension of frequency comb techniques to the mid-infrared spectral region is now under exploration. Versatile mid-infrared frequency comb generators, based on novel laser gain media, nonlinear frequency conversion or microresonators, promise to significantly expand the tree of applications of frequency combs. In particular, novel approaches to molecular spectroscopy in the fingerprint region, with dramatically improved precision, sensitivity, recording time and/or spectral bandwidth may spark off new discoveries in the various fields relevant to molecular sciences.

## 1. Introduction

The mid-infrared spectral region (2-20 µm, 500-5,000 $cm^{-1}$) is a domain of interest to many areas of science and technology. Many molecules undergo strong fundamental tell-tale vibrational transitions in this domain (Figure 1a), which renders mid-infrared spectroscopy a univocal way to identify and quantify molecular species, including isotopologues, in a given environment. It thus provides not only a powerful tool for the understanding of the structure of molecular matter and its governing physical laws, but also for non-intrusive diagnostics of composite systems of physical, chemical or biological interest, in the gas, liquid or solid phase. The mid-infrared region also contains two important windows (3-5 µm and 8-13 µm) in which the atmosphere is relatively transparent. These regions can be exploited to detect small traces of environmental and toxic vapours down to sensitivities of parts-per-billion in a variety of atmospheric, security and industrial applications. The low Rayleigh scattering losses benefit imaging in turbid media and tomography.

Introduced in the late 1990s, laser frequency combs have revolutionized precise measurements of frequency and time. The regular pulse train of a mode-locked femtosecond laser can give rise to a regular comb spectrum of millions of laser modes with a spacing equal to the pulse repetition frequency. Optical frequency combs in the visible and near-infrared regions have enabled the development of new ultra-precise optical atomic clocks and commercially available combs have quickly matured to routine instruments for precision spectroscopy. Frequency combs are now becoming enabling tools for an increasing number of further applications, from the calibration of astronomical spectrographs to molecular spectroscopy. Extensions of frequency comb techniques to new spectral regions from THz frequencies to the extreme ultraviolet are now under exploration and are very likely to trigger new applications of laser frequency combs. In the mid-infrared range, the advent of femtosecond optical combs is not only expected to bring a new set of tools for precision spectroscopy but also to allow to explore spectroscopy in a broader sense. The understanding of molecular structure and dynamics often involves detailed spectral analysis over a broad frequency range. Precise investigations of changes in the composition of a molecular sample over a large dynamic range might also become within reach with novel mid-infrared frequency-comb-based techniques.

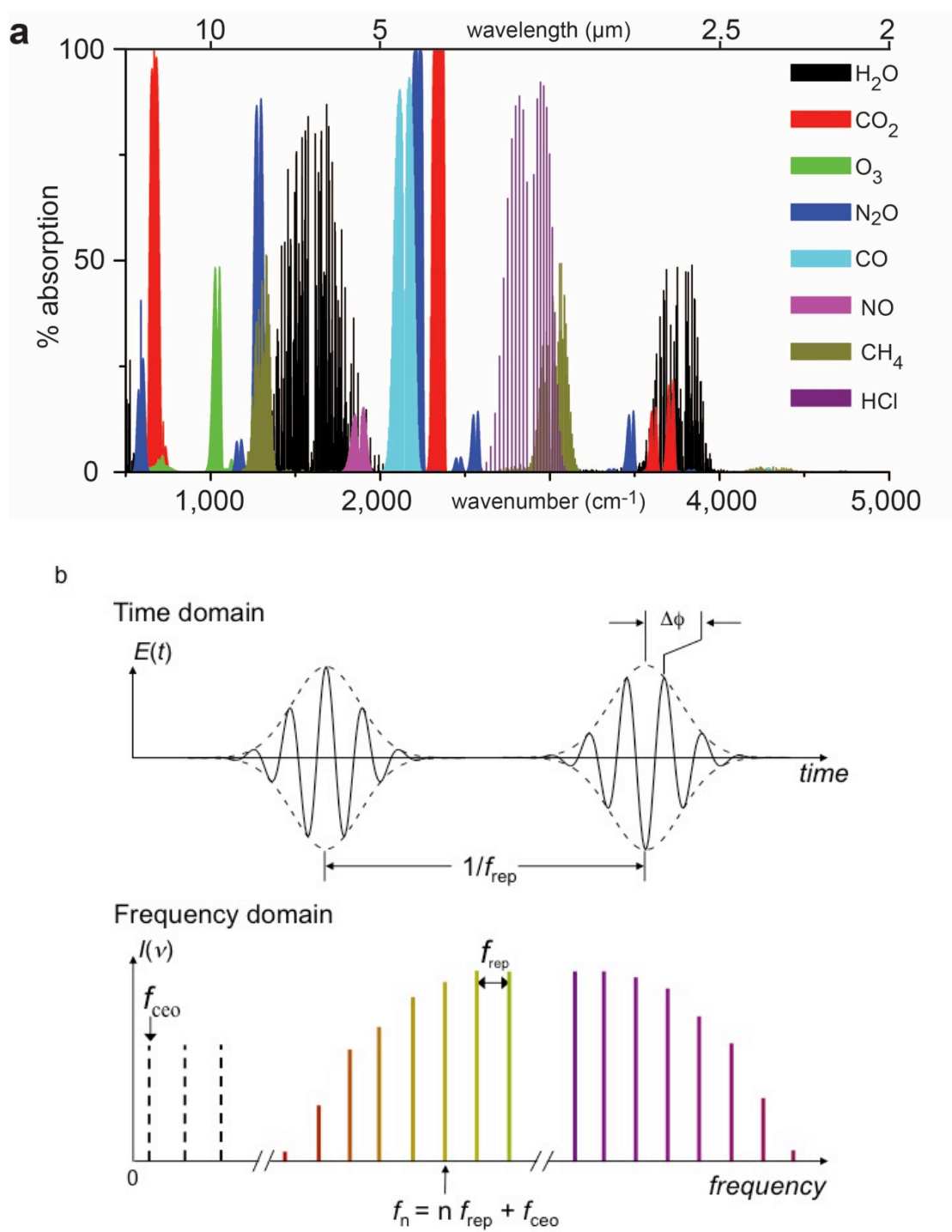

**Figure 1:**
a) Percentage of absorption at line center for the main isotopologue of various molecules in the gas phase in the 20 - 2 μm wavelength region. The calculation assumes a Doppler profile with a pressure of 133 Pa, an absorption path length of 1 cm and a temperature of 296.15 K. The individual line intensities are taken from the 2008 edition[107] of the HITRAN database.
b) Time-domain representation of a frequency comb pulse train and the corresponding frequency-domain picture illustrating the discrete structure of the comb lines.

This review article presents a summary of advances in the emerging field of mid-infrared frequency comb technology and applications. It explicitly focuses on mid-infrared sources, whose comb structure has been established, even if major achievements have also been reported in developing and exploiting other broadband mid-infrared coherent sources. Mid-infrared lasers and photonic technologies have been perfected in five decades of intense research and development, but technical challenges have slowed down the development of convenient instrumentation in this spectral region. In the last few years, however, remarkable progress has been achieved in efficient mid-infrared frequency comb generators. A variety of innovative solutions—including novel laser gain media, non-linear frequency conversion and Kerr comb generators—have been successfully explored. They readily offer a wide choice of comb sources covering vast ranges of repetition frequency and spectral span. Compelling demonstrations in precision spectroscopy and direct frequency comb spectroscopy of molecules in the mid-infrared add enhanced capabilities to spectroscopy and new uses for frequency comb generators. Several new avenues of research are being opened by the introduction of the mid-infrared frequency comb.

## 2. Frequency comb generators in the mid-infrared

### 2.1 Characteristics of a frequency comb

Frequency combs[1-6] are traditionally generated by means of mode-locked lasers: Every time the pulse circulating in the laser cavity reaches the output coupler, a fraction of its energy is coupled out of the laser. The emitted pulse train therefore has a strictly periodic envelope function, with a repetition frequency given by the inverse round-trip time of the pulse in the laser cavity $f_\text{rep} = L/v_\text{g}$, where $v_\text{g}$ is the group velocity of the light in the cavity, and $L$ the round-trip length of the latter. However, due to dispersion in the laser resonator, the carrier wave of the pulse propagates at a phase velocity $v_\text{p}$ different, in general, from $v_\text{g}$. Therefore, the electric field of the out-coupled pulses is shifted with respect to the pulse envelope by a constant amount $\Delta\varphi$ from pulse to pulse. This implies that the electric field is not strictly periodic in time. Nevertheless, the Fourier spectrum of the electric field is composed of a discrete set of equidistantly spaced, sharp frequency components $\nu_n$ obeying the simple relation

$$\nu_n = n \cdot f_\text{rep} + f_\text{ceo}$$

in which the carrier-envelope offset frequency is given by $f_\text{ceo} = \frac{\Delta\varphi}{2\pi} f_\text{rep}$, and $n$ is a large integer (Figure 1b).

### 2.2 Mode-locked lasers

Generation of mid-infrared radiation with such spectral properties can be achieved along several routes. Similar to the established sources in the near-infrared, mode-locked lasers can be realized in the mid-infrared (Figure 2a). Transition metals such as chromium or iron provide a broad gain bandwidth in the mid-infrared region if doped into crystals of II-VI compounds[7-10]. Electronic transitions in these laser-active metals are strongly coupled to phonons in the host crystals' lattice, which leads to significant homogeneous broadening, often in the range of 30% of the centre frequency and more. Chalcogenide hosts such as ZnSe and ZnS are particularly well suited, since they support this "vibronic" broadening, but undesirable non-radiative multi-phonon relaxation is still sufficiently suppressed—even at room temperature—due to their low maximum phonon energy.

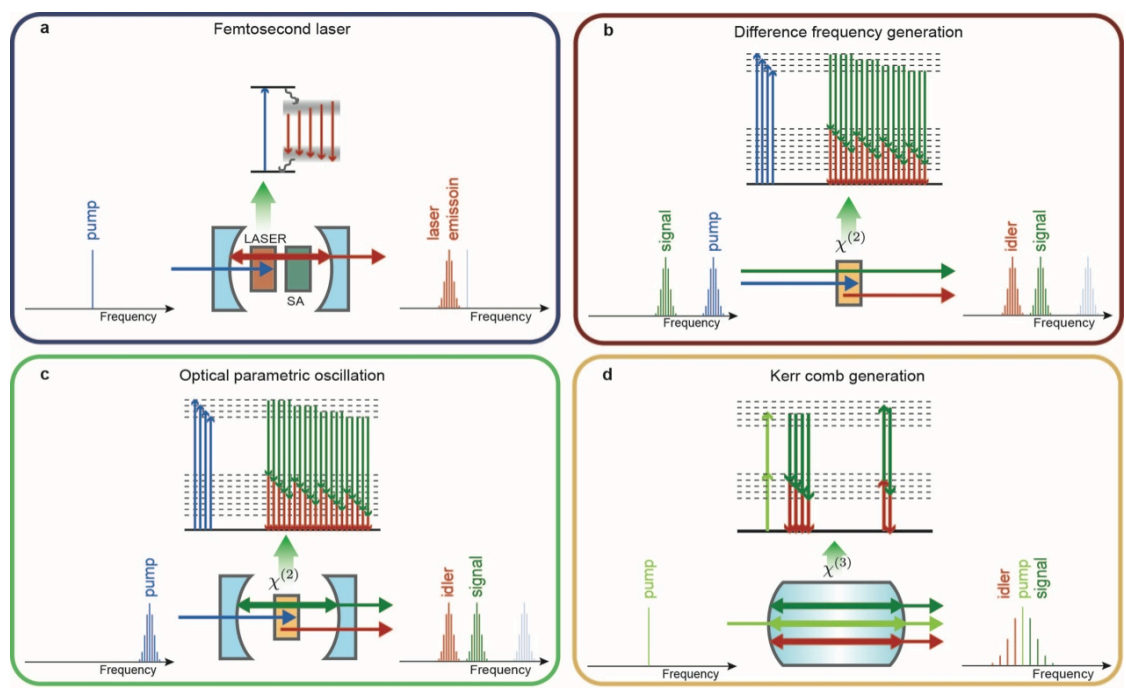

**Figure 2 : Approaches to frequency comb generation in the mid-infrared.**
a) Femtosecond mid-infrared laser. A laser-active medium with broad gain is inverted by a strong optical pump field. In a resonator that additionally holds an (effective) saturable absorber (SA), a short laser pulse builds up that circulates in the cavity. The spectrum of the out-coupled laser light constitutes a frequency comb.
b) Difference frequency generation. In a medium with a second-order nonlinearity $\chi^{(2)}$, nonlinear frequency mixing of the multitude of modes of two "signal" and "pump" combs leads to the generation of various difference frequencies, forming an "idler" comb in the mid-infrared. Alternatively, either signal or pump can also be a single continuous-wave laser.
c) Optical parametric oscillation. A $\chi^{(2)}$-nonlinear medium placed in an optical resonator provides gain to the cavity's "signal" modes if pumped by a strong pump beam. If this gain exceeds the signal modes' losses, coherent oscillation of signal modes, seeded only by the intracavity quantum fluctuations, sets in. Due to energy conservation, the desired mid-infrared idler modes are concomitantly generated.
d) Kerr comb generation. A dielectric resonator made of a $\chi^{(3)}$ nonlinear material is pumped by a mid-infrared continuous wave source. Signal and idler sidebands to the pump are generated in the modes of the resonator adjacent to the pump mode via degenerate four-wave-mixing. Due to the high quality factor of the modes, this process can be very effective and cascade to a large number of sideband modes. In addition, non-degenerate four-wave-mixing can lock the phases of all generated sidebands.

Using semiconductor saturable absorber mirrors (SESAMs) or the Kerr lens effect, mode-locked operation, at typical 100 MHz repetition rates is now achieved in $Cr^{2+}$:ZnSe[11-13] and $Cr^{2+}$:ZnS[14] lasers, emitting relatively wide (> 300 cm$^{-1}$)

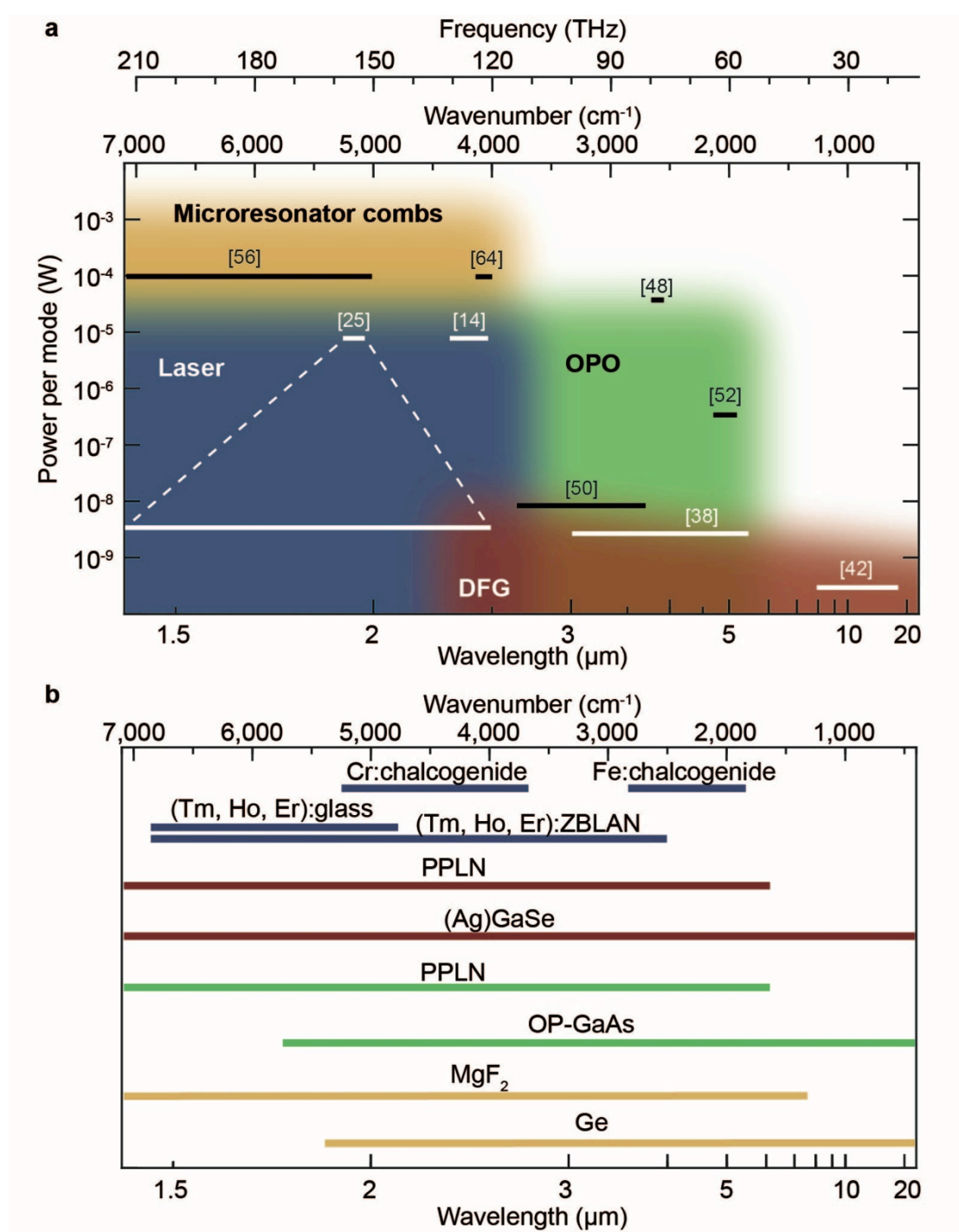

**Figure 3: Spectral coverage of different approaches and materials.**
a) Overview of the spectral regions and power range accessible with mid-infrared femtosecond lasers (blue), difference-frequency generation (red), optical parametric oscillators (green) and micro-resonator-based Kerr combs (yellow). Black and white bars represent a selection of realized comb sources. Due to an often strong variation of the power per mode across the generated spectrum, the following representation was adapted: The bars' width represent the spectral span over which the respective comb has at least the power per mode indicated by its position on the ordinate. Note that many sources are additionally wavelength-tunable (not shown).
b) Potential spectral coverage of selected materials if employed in mid-infrared comb generators (color code as in a)). Note that the range indicated for the rare-earth ions covers several different transition bands in each of the ions spectra around 2.5 μm wavelength, with powers on the order of 10 μW per mode (Figures 3 and 4a). Their comb structure has been indirectly established by their successful application in dual-comb spectroscopy[15] (see below). Further engineering could give access to the full, ~2,000 cm$^{-1}$ wide gain bandwidth of chromium ions at 2.5 μm. Within the 4-5 μm region, tuneable lasing has been demonstrated in Fe$^{2+}$-doped crystals[16-18]. Such lasers, if mode-locked, could directly provide frequency comb sources deep in the mid-infrared. Furthermore, thulium (Tm), holmium (Ho) and also erbium (Er) ions have several gain bands in the mid-infrared region[19] (Figure 3b). Doped into silica fibre, Tm- fibre lasers emit around 1.9 μm, and mode-locking has been achieved with a variety of methods[20-24]. Phillips et al. have recently demonstrated a spectral coverage of 190 cm$^{-1}$ with such a laser[25] (Figure 3). In addition, powerful amplifiers based on doped fibres[25,26] or crystals[27] can boost the power of these combs to the Watt level, facilitating also their additional spectral broadening in non-linear fibre[25]. Such fibre-based systems may quickly evolve into turn-key alignment-free instruments.

## 2.3 Difference frequency generators

Beyond 3 μm wavelength, laser sources have remained scarce. Quantum cascade lasers[28] can be tailored to operate at nearly arbitrary wavelengths, and recent experiments indicate the principle possibility of mode-locking[29,30]. However, the comb properties of these rather narrow (<50 cm$^{-1}$) emission spectra have not yet been put to test. In most cases, a different route is taken to access the region beyond 3 μm: Non-linear optical effects are harnessed to transfer electromagnetic energy from the visible or near-infrared domain into the mid-infrared (Figure 2b,c).

For example, all modes of a near-infrared frequency comb can simultaneously be subjected to difference frequency generation[31] (DFG). When mixed with a continuous-wave laser of frequency $v_{cw}$, the modes of a comb with frequencies $v_n = n \cdot f_{rep} + f_{ceo}$ can correspondingly be transferred to $v_n^{DFG} = |v_n - v_{cw}|$, lying in the mid-infrared for an appropriately chosen $v_{cw}$. The efficiency of this process depends not only on the strength of the optical non-linearity, equally important is the ability to achieve phase-matching—that is, constructive interference of the radiation generated at different depths in the nonlinear crystal—over the wide necessary bandwidth. For broadband conversion, a widespread approach is to use periodically poled crystals, in which the sign of the nonlinear coefficient is reversed just at the depth where the generated waves would start to oscillate out of phase. In that manner, coherent build-up of the difference frequency all along the employed crystal is supported. However, even in this case, imperfect phase-matching is often the limiting factor for the achieved bandwidth, as the poling period is optimized for a part of the comb spectrum only. This approach[32,33] typically achieves 200 cm$^{-1}$ spectral width and powers on the order of nanowatts per mode (Figure 3). Difference frequency generation between two synchronized combs has also been poorly investigated[34], due to its complexity and cost. Alternatively, difference frequencies can be generated between the different teeth of one and the same comb. As in this case $v_{n,m}^{DFG} = |n - m| \cdot f_{rep}$, with both $n$ and $m$ integer, the resulting comb's carrier-envelope offset frequency is fixed to zero, simplifying the control of the modes' absolute frequency. Using, as before, periodically poled lithium niobate (ppLN) as a nonlinear medium, spectra between 3 and 5 μm, more than 500 cm$^{-1}$ wide and with powers of up to 10 nW per comb mode have been generated[35-38]. At about 5 μm wavelength, absorption in lithium niobate sets a limit to how far this approach can be pushed into the mid-infrared. Other materials[31] must be used to access longer wavelengths (Figure 3b). For example, 300 cm$^{-1}$ wide spectra, tunable between 3 and 17 μm have been generated in gallium selenide with up to ~50 nW power per comb tooth[39-43] (Figure 4b).

## 2.4 Optical parametric oscillators

A key challenge in sources based on nonlinear optics is the efficiency of the desired photon conversion process. It often remains well below 1%, even if there are promising exceptions[36]. One route to boost efficiency is the use of optical parametric oscillators (OPO, Figure 2c)[44-46]. If pumped by a femtosecond laser, many longitudinal "signal" modes of the OPO's resonator can simultaneously experience gain exceeding the threshold for parametric oscillation. Of course, achieving oscillation of many signal and idler modes requires, in addition to phase matching, management of the resonator's dispersion to ensure that the equidistant signal modes coincide with the resonator's modes across the desired wide bandwidth. In a time-domain picture, this corresponds to the requirement that the signal modes' power remains concentrated in a short pulse, passing the non-linear medium synchronously with the pump pulses. With ppLN-based OPOs pumped by compact solid-state[47] or powerful fibre-laser based near-infrared combs[48], wide (>100 cm$^{-1}$) spectra with up 30 µW per mode, tuneable all over a ~2-4µm wavelength range have been demonstrated.

While the signal and idler combs inherit the mode spacing $f_{\text{rep}}$ from the pumping comb, the carrier-envelope-offset frequency of these combs is only subject to the condition $f_{\text{ceo}}^{\text{signal}} + f_{\text{ceo}}^{\text{idler}} - f_{\text{ceo}}^{\text{pump}} = n \cdot f_{\text{rep}}$. For a fully stabilized pump source ($f_{\text{rep}}$ and $f_{\text{ceo}}^{\text{pump}}$ fixed), one degree of freedom must still be controlled. In many recent implementations of mid-infrared OPO sources, this is facilitated by the generation of optical sum- and second harmonic frequencies occurring concomitantly with the actual OPO action in the nonlinear crystal[46-48], which allows to determine and control $f_{\text{ceo}}^{\text{signal}}$ and therefore also $f_{\text{ceo}}^{\text{idler}}$. In the special case of a degenerate OPO[49], in which signal and idler combs merge into one comb of subharmonic frequencies, the additional constraint $f_{\text{ceo}}^{\text{signal}} = f_{\text{ceo}}^{\text{idler}}$ allows only the values $f_{\text{ceo}}^{\text{pump}}/2$ and ($f_{\text{ceo}}^{\text{pump}} + f_{\text{rep}}$)/2 for the subharmonic comb's carrier-envelope-offset frequency, simplifying stabilization further. In addition, such a degenerate OPO enables the generation of ultrabroad mid-infrared spectra spanning, for example, 900 cm$^{-1}$ around 3 µm in a recent demonstration[50] with an impressive power of 200 nW per mode. Another noteworthy route is the use of orientation-patterned GaAs as nonlinear medium, which allows accessing longer wavelengths, as its transparency range extends deeper into the mid-infrared[51,52].

## 2.5 Microresonator-based Kerr combs

Recently, a special form of parametric oscillation (sometimes referred to as hyperparametric oscillations[53]) has been reported to enable frequency comb generation in optical microresonators[54] (Figure 2d). These dielectric resonators consist of an essentially toroidal ring of sub-millimetric diameter, which host very high quality optical whispering-gallery modes (WGMs). If populated with a strong pump field, the material's *third*-order non-linearity $\chi^{(3)}$ can lead to four-wave mixing (FWM). For a single-frequency continuous-wave pump, this implies the generation of a pair of signal and idler photons, whose combined energy corresponds to the energy of *two* pump photons, $h\nu_{\text{signal}} + h\nu_{\text{idler}} = 2h\nu_{\text{pump}}$. Due to the high finesse of the resonators (up to $10^7$), the threshold for this process can be particularly low if all involved fields are resonant with cavity modes. This is possible if dispersion and non-linear phase shifts due to self- and cross-phase modulation are appropriately balanced[55]. Remarkably, a single-frequency pump laser can give rise to not only one signal-idler pair, but a massive cascade of signal and idler sidebands, which additionally become mutually phase-locked by non-degenerate FWM which only require $h\nu_{\text{signal}} + h\nu_{\text{idler}} = h\nu'_{\text{signal}} + h\nu'_{\text{idler}}$. In that manner, "Kerr" combs can be generated that can cover an entire octave[54,56,57], extending already into the mid-infrared if pumped by a strong near-infrared laser. As an aside we note that a similar mechanism is at work in the generation of mid-infrared "supercontinua", reported to occur in waveguides made of infrared-enabled materials. However, present experiments[58-61] have not yet shown that the coherence of the resulting spectra is not destroyed by competing nonlinearities.

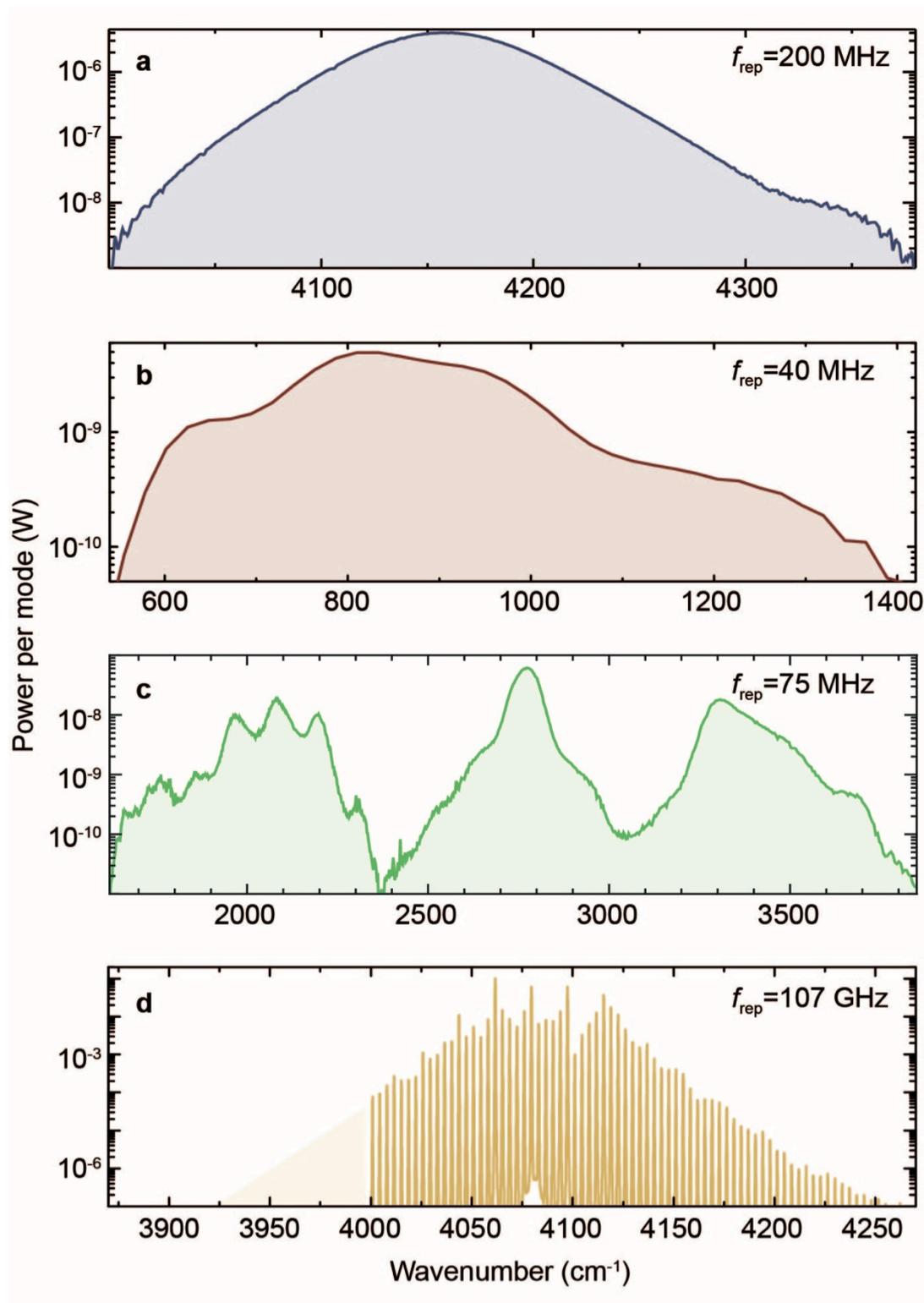

**Figure 4: Typical mid-infrared spectra obtained with different approaches.**
a) Femtosecond Cr:ZnSe laser (Ref [11]),
b) Difference frequency generation in GaSe (Ref [42]),
c) Optical parametric oscillation in GaAs (Ref [51]),
d) Kerr comb generation in a $MgF_2$ microresonator (Ref [64]).
Note that only in d) the individual comb components could be resolved by the employed spectrograph. The shaded area in this panel indicates expected spectral components outside the spectrograph's measurement range.

Resonators made from crystalline materials such as calcium and magnesium fluoride support ultra-high-Q WGM still deep in the mid-infrared ($\lambda \lesssim 7\mu\text{m}$) and simultaneously exhibit the anomalous dispersion required for efficient and low-noise comb generation[62,63]. If pumped at 2.45 µm, a Kerr comb covering the range from 2.35 µm to beyond 2.5 µm (at the -60 dB level) has been reported[64] (Figure 4d). A powerful pump source, and the large mode spacing of 107 GHz corresponding to the round-trip along the resonator's perimeter enable very high power per mode up to the milliwatt scale (Figures 3 and 4d). This novel approach thus enables simple and compact mid-infrared comb generators, with yet unexplored opportunities with different resonator materials (Si, Ge, InP) or pump sources. Finally, the large mode spacing could be beneficial for a number of emerging applications in spectroscopy and metrology as described below.

## 3. Applications

Applications of mid-infrared frequency combs have logically first concentrated on precision spectroscopy of molecules. Recent experiments now demonstrate an intriguing potential for the rapid and sensitive acquisition of molecular spectra over a vast spectral span. Foreseen applications might push even further the frontiers of spectroscopy and dynamics with tools including coherent control or non-linear phenomena.

### 3.1 Frequency combs as spectroscopic calibration tools

Frequency combs have been initially conceived for precision spectroscopy of the simple hydrogen atom and they are now commonly used in the visible and near-infrared spectral regions as tools for precise frequency measurements of atomic resonances and some molecular rovibronic transitions. The mid-infrared spectral region offers complementary opportunities due to the presence of strong rovibrational transitions of a number of molecules (Figure 1a). Accurate frequency measurements in the mid-infrared provide a better understanding of the energy levels of a molecule. As frequency combs can conveniently link optical and microwave frequencies they provide the long missing clockwork for optical clocks. In the mid-infrared region, they have enabled the establishment of molecular optical clocks, using e.g. the hyperfine $F_2^{(2)}$ optical transition[35,37] of the P(7) line of the $\nu_3$ band in methane as a pendulum with an instability of 1.2 10$^{-13}$ at 1 s. This allows the comparison between a mid-infrared clock and other optical clocks.

Comb-enabled accurate frequency measurements (Figure 5a) in the mid-infrared also allow new tests of fundamental physical laws and improved determination of fundamental constants, as well as sensitive limits for their possible slow variations: For instance, the frequency comparison of a rovibrational transition in SF$_6$ (Fig. 5b) around 10 µm -accessed though spatial two-photon Ramsey fringes- to a cesium microwave atomic clock by means of a frequency comb and an optical link[65] link has set[66] an upper limit on a possible small variation of the proton-to-electron mass ratio. Such model-free results are derived from an absolute frequency measurement within an impressive 2.8 $10^{-14}$ fractional uncertainty. They complement astronomical observations, where the measurements are effectively separated in time by several gigayears.
Furthermore, the Boltzmann constant is presently determined by acoustic gas thermometry within an inaccuracy of 1.7 $10^{-6}$. An accurate measurement of the Doppler profile of a well-isolated absorption line of a molecular gas in thermal equilibrium in a cell, such as the $\nu_2$ saQ(6,3) rovibrational line of $^{14}NH_3$ around 10 µm, has been explored[67] for a few years as a potential way to measure the Boltzmann constant by optical means. This endeavor requires, in addition to in-depth lineshape modeling and control of the systematic effects, accurate control of the frequency of the laser probing the absorption profile. As a last example, measuring small differences between infrared transition frequencies of the two enantiomers of a chiral molecule might evidence[68] parity violation due to weak interaction. However the latest predictions estimate such relative differences to be as small as 8 x $10^{-17}$ for the CHFClBr molecule and of the order of $10^{-13}$-$10^{-14}$ for oxorhenium complexes around 10 µm. The latter sensitivity, although technically challenging with such complex molecules, might become within reach with the state-of-art ultra-high resolution apparatus used in [66].

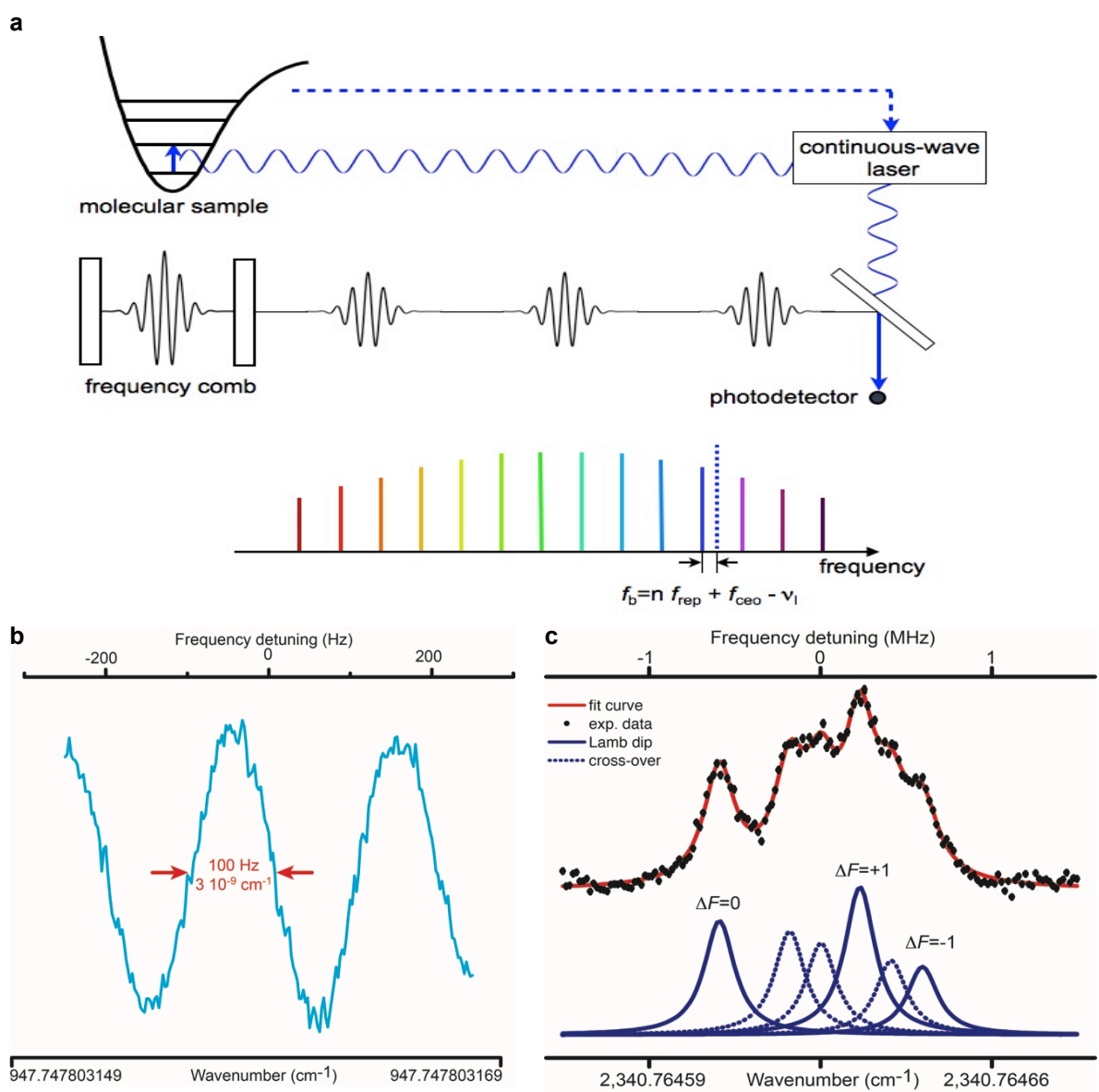

**Figure 5: The frequency comb as a frequency ruler.**
a) Laser frequency combs phase-coherently compare optical and microwave frequencies in a single step. The frequency of any comb line can be calculated from the two radio-frequencies $f_{rep}$ and $f_{ceo}$, that can be compared or phase-locked to a microwave reference, together with the integer mode number n.
Optical frequency measurements are often aimed at determining the absolute frequency of an atomic or molecular transition interrogated by a single-frequency continuous-wave laser. By heterodyning the continuous-wave laser against a nearby optical comb line, the frequency of the continuous-wave laser, $\nu_l$, is measured: the resulting beat note $f_b$ can be counted with standard radio-frequency equipment.
b) Comb-assisted ultra-high resolution mid-infrared spectroscopy[66]. A spatial Ramsey interferometer resolves within 100 Hz the P(4)E$^0$ rovibrational transition in the 2$\nu_3$ band of SF$_6$. The 10 μm CO$_2$ laser used to interrogate the two-photon resonance in a molecular beam, is referenced[70] via a frequency comb and an optical link to a cesium fountain clock. The frequency measurement is achieved within an uncertainty of 0.79 Hz (2.6 10$^{-11}$ cm$^{-1}$).
c) Comb-assisted Doppler-free ultrasensitive mid-infrared spectroscopy[74]. Saturated-absorption cavity ring-down spectroscopy with a tunable continuous-wave laser is a technique that combines sensitivity, high resolution and frequency accuracy through frequency comb referencing. The hyperfine triplet –and related crossovers- of the R(0) transition of the $\nu_3$ band of the rare isotopologue $^{17}$O$^{12}$C$^{16}$O around 4.27 μm is observed at at a partial pressure of 150 μPa. With a Rb/GPS-disciplined frequency comb, its line positions are measured within 1.0 10$^{-7}$ cm$^{-1}$.

For precision spectroscopy in the mid-infrared, the frequency comb is used (Figure 5a) as a frequency ruler, similar to the way it is used in the near-infrared or visible ranges. The unknown absolute frequency $\nu_l$ of a continuous-wave laser is determined by creating a beat note $f_b$ with the nearest comb mode: $\nu_l$ = n $f_{rep}$ + $f_{ceo}$ ± $f_b$. The correct sign of $f_b$ may be determined by a small change in either of the frequencies, and the mode number n may be determined by a coarse measurement of $\nu_l$, for example, with a wavemeter. Although referencing the continuous-wave laser to a mid-infrared frequency ruler has been demonstrated, such experiments have by far not yet reached the performance of their visible counterparts. Frequency combs have been produced by difference frequency generation for direct referencing of methane-stabilized HeNe lasers[35,37] at 3.39µm or difference-frequency generation continuous-wave sources[69] at 3 µm. The main drawback of this approach is that both the comb and the probing laser emit in the mid-infrared region, where low-noise high-speed detectors, and more generally advanced photonics technology, are difficult to be found. Therefore up-conversion of the infrared light from the continuous-wave laser has often been a preferred alternative. Sum-frequency generation between the continuous-wave laser with the near-infrared comb produces a frequency-shifted comb that may still overlap the initial comb. The resulting phase-coherent beating signals can be used to determine the continuous-wave laser's frequency (modulo the comb spacing), independently of the value of the carrier-envelope offset frequency of the comb. Such an approach has been implemented for e.g. Doppler-free two-photon spectroscopy of $SF_6$ and saturated-absorption spectroscopy of formic acid with $CO_2$ lasers around 9-10 µm[70,71] or Doppler-limited spectroscopy of carbon dioxide with quantum cascade lasers at 4.3 µm[72]. The metrological link to a near-infrared frequency comb may also be realized[73] by sum frequency generation between two continuous-wave lasers, as demonstrated e.g. with a 4.4 µm quantum cascade laser mixed to a 1.064 µm Nd:YAG laser. Alternatively, when the infrared radiation of the continuous-wave probing laser is produced by nonlinear frequency down-conversion of near-infrared lasers, the near-infrared continuous-wave laser may be referenced to a near-infrared frequency comb, as already implemented in schemes involving difference frequency generation[74,75] (Fig. 5c) and optical parametric conversion[76].

## 3.2 Direct frequency comb spectroscopy

Direct frequency comb spectroscopy, a denomination that first appeared in Ref.[77], has been the first application[78,79] of frequency combs, in the mid-seventies with picosecond mode-locked dye lasers. Direct frequency comb spectroscopy of atomic resonances by Ramsey-like excitation with a coherent train of multiple light pulses and Doppler-free two-photon excitation has been long used, mostly in the visible and ultraviolet region. The fluorescence from the excited sample is monitored while either the repetition frequency $f_{rep}$ or the carrier-envelope offset $f_{ceo}$ is tuned. The comb is then used in a similar manner as a continuous-wave laser, with the advantage that the high peak intensity of pulsed lasers facilitates efficient nonlinear conversion into frequency regions where continuous-wave lasers are not available.

In recent years, novel techniques[80-85] have been developed in which the comb directly interrogates a vast number of transitions of an absorbing sample. To date direct frequency comb absorption spectroscopy of molecules in the gas phase has been the most prominent application of mid-infrared frequency combs. The frequency comb indeed has to emit in the region of the absorption lines of interest. Therefore the mid-infrared domain, with its strong molecular fingerprints, attracts considerable interest. Direct absorption frequency comb spectroscopy may result in short measurement time, high sensitivity and high accuracy over a broad spectral bandwidth. The spectrum of the comb is frequency-selectively attenuated and phase-shifted by the molecular resonances and needs a spectrometer to be analysed. As dispersers associated with detector arrays are not efficient and conveniently available in the mid-infrared region, the preferred approach has been Fourier transform spectroscopy[86]. Fourier spectrometers record the spectroscopic signal on a single photodetector and therefore overcome the dispersive-related issues. Two different implementations of Fourier transform spectrometers have been successfully reported: Michelson-based and dual-comb spectroscopies. They present distinct advantages but use the same physical principle (Figure 6a,b): the frequencies of the mid-infrared comb are too high (15-150 THz) to be directly counted. Therefore the optical spectrum is down-converted into a lower frequency region, where it becomes accessible to digital signal processing. Frequency comb Fourier transform spectroscopy is actually a time domain technique in which the pulse train of a comb is interferometrically sampled, akin to an optical sampling oscilloscope, by a second pulse train of different repetition frequency. The latter may be produced by the varying optical delay of a Michelson interferometer or by a second frequency comb source.

In Michelson-based frequency comb Fourier transform spectroscopy (Figure 6c), the pulse train of a frequency comb with a pulse repetition frequency frep is analyzed by a Michelson interferometer that records interference versus path difference 2vt (t: time, v: constant velocity of the moving mirror).  The comb traveling along the fixed mirror arm keeps its native line-spacing $f_{rep}$. The comb in the interferometer moving arm has a Doppler-shifted repetition frequency equal to $f_{rep}(1 - 2v/c)$. The two interfering frequency combs exiting from the interferometer interrogate the absorbing sample and beat on a photodetector. Due to the limited speed of the moving mirror, the optical frequencies are down-converted in the acoustic range of <100 kHz. Simply analyzing a mid-infrared frequency comb light source —as any broad spectral bandwidth coherent source— with a Michelson-based Fourier transform spectrometer dramatically reduces the measurement time (or improves the signal-to-noise ratio) due to the high spectral radiance of the coherent source. This has been demonstrated with a Cr:ZnSe mode-locked laser[11] around 2.4 µm and an optical parametric oscillator-based frequency comb[87] tunable between 2.8 µm and 4.8 µm with a span ranging between 50 and 400 cm$^{-1}$. The sensitivity for weak absorption may be much enhanced when the effective absorption path-length is increased by means of a multi-pass cell or a high finesse resonator. In the latter scheme, the optical cavity is matched to the laser resonator so that it is simultaneously resonant for many comb lines. Therefore dispersion-managed cavity mirrors, control of the two degrees of freedom of the comb for coherent addition of the pulses circulating inside the resonator, as well as tight-locking of the comb to the cavity to avoid frequency-to-amplitude noise conversion, are required. The strong intensity of the rovibrational mid-infrared lines associated with a sensitivity enhancement in the cavity of several thousands allows[88] to reach detection levels as low as one part per billion in 1s measurement time for the $H_2O_2$ molecule around 3.75 μm (Figure 6d). Another intriguing application explores[89] the potential of broadband near-field microscopy with a frequency comb produced by difference frequency generation around 1,000 $cm^{-1}$ to map interferograms of the near-field interaction between a metal tip and polar materials within a few seconds per tip-illuminated spot.

While Michelson-based Fourier transform spectroscopy is a technique that has been mastered for more than 40 years and as such, the replacement of the traditional incoherent light source by a frequency comb does not create major instrumental difficulties, dual-comb spectroscopy is in its infancy and its full potential is still to be realized.
In dual-comb spectroscopy (Figure 6e), two frequency combs, with slightly different line spacing, are transmitted through the cell and are heterodyned, yielding a down-converted radio-frequency comb containing information on the absorption experienced by the lines of the combs. Other implementations allow a single comb to interrogate the sample and to also measure the dispersion induced by the sample. The static scheme of dual-comb spectroscopy overcomes the shortcomings induced by the moving mirror of the Michelson interferometer. The resolution is only limited by the line spacing of the combs and the measurement time. Interleaving several spectra may improve the resolution down to the comb intrinsic linewidth. Moreover, the radio-frequency domain up to $f_{rep}/2$ may be covered, being only restricted by aliasing. In comparison to Michelson-based Fourier transform spectroscopy, recording times may be dramatically reduced. References [82,92,93,94] (and references therein) provide detailed information on the practical implementation of a dual-comb spectrometer.
Low-resolution, proof-of-principle demonstrations of dual-comb spectroscopy with difference frequency generators have been carried out in the 10 μm region since 2004, probing gaseous[39,40] and solid-state[90] samples. Although limited in terms of optical power and stability of the combs, they highlighted one promising characteristic of dual-comb spectroscopy: very short measurement time essentially enabled by the absence of moving parts in the spectrometer. Frequency-comb-based scanning near-field optical microscopy has also been explored similarly[91]. The lack of efficient mid-infrared frequency comb sources and the difficulty of synchronizing the pulse trains of two combs with interferometric precision have hindered the fast development of mid-infrared dual-comb spectroscopy and most reports[82,92-94] have thus privileged the near-infrared domain. Stimulated Raman dual-comb spectroscopy[95] is an alternative technique to access fundamental transitions with near-infrared combs. However in recent times, a proof-of-principle demonstration[15] with unstabilized Cr:ZnSe mode-locked lasers around 2.4 μm has enabled the recording of the rovibrational spectrum of acetylene between 3,960 and 4,220 $cm^{-1}$ within 10 μs at 12 GHz resolution. A difference-frequency generation dual-comb set-up has shown[33] (Figure 6,f) around 3.4 μm a high accuracy of $10^{-5}$ $cm^{-1}$ in the measurement of the line positions of the Doppler-broadened transitions of the $\nu_3$ band of methane Simplification of the laser systems for dual-comb spectroscopy is also highly desirable: an intriguing option is offered by a 3.3 μm femtosecond optical parametric oscillator emitting[96] two asynchronous trains of pulses, with no significant nonlinear coupling between the two channels.

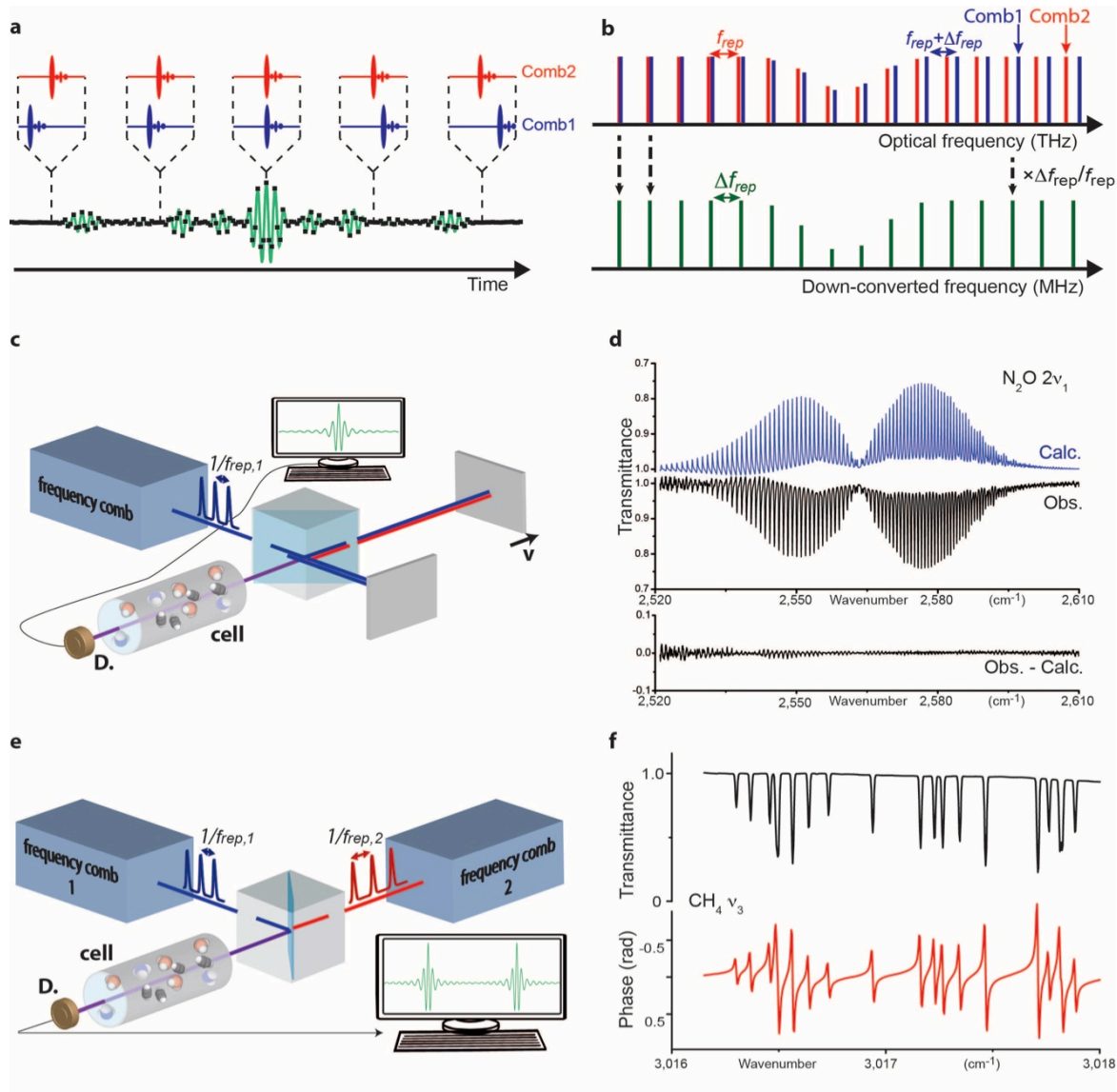

**Figure 6: Frequency comb Fourier transform spectroscopy.**
(a) Time domain picture of the physical principle of Fourier transform spectroscopy. The pulse train of comb 1 walks through the pulses from comb 2 to generate a cross-correlation interferogram $I(t)$.
(b) Frequency domain picture of the physical principle of Fourier transform spectroscopy. Due to interference between pairs of optical comb lines, the optical absorption spectrum is effectively mapped into the acoustic or radio-frequency region, where it becomes accessible to fast digital signal processing. The frequencies of the optical spectrum are down-converted to $\Delta f_{ceo} + n \Delta f_{rep}$, where $\Delta f_{ceo}$ is the difference of carrier-envelope slippage frequencies of the two combs, n an integer and $\Delta f_{rep}$ the difference of comb repetition frequencies. With a symmetric Michelson interferometer, $\Delta f_{ceo}=0$
(c) Michelson-based frequency comb Fourier transform spectroscopy.
(d) Rotationally-resolved experimental (obs.) and computed (calc.) spectra[88] of the $2\nu_1$ overtone band of $N_2O$, around 3.9 µm, by cavity-enhanced frequency comb Fourier transform spectroscopy at a resolution of $28 \cdot 10^{-3}$ cm$^{-1}$. The enhancement resonator with a finesse exceeding 3800 contains 0.8 ppm of nitrous oxide at 1013 hPa of nitrogen.
(e) Dual-comb spectroscopy.
(f) Portion[33] of absorption and dispersion dual-comb spectra showing the Q(8)-Q(4) multiplets of the $\nu_3$ band of methane. In ref. [33], each spectrum of methane (pressure: 27 Pa, absorption length: 28 cm) spans 33 cm$^{-1}$ and is measured, with a signal-to-noise ratio of 3500, within 13 minutes at $3 \cdot 10^{-3}$ cm$^{-1}$ resolution. In total, the positions of 132 rovibrational transitions, spanning over 150 cm$^{-1}$, are determined within an uncertainty of 300 kHz.

Michelson-based frequency comb Fourier transform spectroscopy and dual-comb spectroscopy share a number of advantages over traditional spectrometric instruments: as with any Fourier transform spectrometer, the spectral span is only limited by the source and single detector spectral bandwidths. The multiplex advantage grants overall consistency of the simultaneously measured spectral elements that may prove crucial when e.g. investigating transient phenomena. The spatial directionality of laser sources permits remote measurements and microscopic focusing. The high spectral radiance of frequency comb sources provides much-improved signal to noise ratios and reduced measurement times. The comb structure of the spectrum dramatically improves the detection sensitivity via optimal use of cavity-enhanced techniques. Moreover, resolving the comb lines allows for self-calibration of the frequency scale of the spectra. Analyzing the comb light with a Michelson interferometer readily offers robust, fast and sensitive multiplex spectrometers and the already high quality of the achievements guarantees immediate opportunities for many applications, e.g. in trace gas detection of multiple species. Dual-comb spectroscopy is more technically challenging and certainly still requires considerable efforts to express its full potential. However, recording of Doppler-limited 300 $cm^{-1}$-broad mid-infrared spectra within 100 µs is clearly within reach—as are resolutions only limited by the comb linewidth, fully justifying the efforts. The absence of moving parts in dual-comb spectroscopy indeed overcomes the speed and resolution limitations of the Michelson-based approach. Further, compact high-resolution instruments may be designed and one may even envision a chip-size dual-comb spectrometer based on microresonators for real-time spectroscopy in the liquid phase.

The combination of frequency combs with other advanced tools of laser science, nonlinear optics, photonics, and electronic signal processing may vastly enhance the range and capabilities of molecular laser spectroscopy. Molecular physics with laser frequency combs is however still in the early stages of its development. Both techniques of frequency comb Fourier transform spectroscopy may be combined with e.g. Doppler-free measurement schemes, hyperspectral imaging, microscopy, temporal resolution or selective molecular detection techniques, while keeping their already demonstrated capabilities. More interestingly, since laser frequency combs involve intense ultrashort laser pulses, nonlinear interactions can additionally be harnessed and, by the control of the phase of the electric field of ultrashort laser pulses and the promises of line-by-line pulse shaping, the combination of coherent control over quantum mechanical processes and high-sensitivity broadband frequency comb spectroscopy techniques might be envisioned. Frequency comb Fourier transform spectroscopy might establish the basis of ground-breaking spectroscopic tools and open up new insights in the understanding of the structure of matter as well as new horizons in advanced diagnostics instruments, for instance in chemistry or biomedicine.

### 3.3 Future applications

Beyond enabling advances in molecular spectroscopy, mid-infrared frequency combs are foreseen to lend tremendous value to existing applications and they may be envisioned to strongly impact emerging or unexpected fields for many areas of research, similarly to their rapidly evolving expansion in the visible and near-infrared spectral domains.

For instance, line-by-line pulse shaping[97] of mid-infrared frequency combs would allow for coherent control of e.g. molecular vibrational excitation in the ground electronic state, or elementary excitations in condensed matter systems. The traditional shaping devices are either not transparent or show very low diffraction efficiency in this wavelength region, which obstructs the direct modulation of the electric field transients. One approach may be to shape the signal beam of an OPO-based frequency comb.

Frequency combs in the near-infrared and visible regions have recently enabled great strides[98,99] in the improvement of the calibration of ground-based astronomical spectrographs. The atmospheric windows in the mid-infrared benefit the radial velocity technique[100] for detection of high red-shift galaxies and exoplanets and the spectroscopic characterization of their atmospheres. Several existing or planned mid-infrared cross-dispersed echelle spectrographs at large telescope facilities, like the CRyogenic high-resolution InfraRed Echelle Spectrograph[101] at the Very Large Telescope, may profit from calibration by mid-infrared frequency combs with a large mode spacing.

Mid-infrared frequency combs can serve as pump sources to further extend the spectral territories covered by frequency comb techniques. Femtosecond thulium fiber lasers are already used[51] to synchronously pump OPO spanning the 3-6 µm range. Mid-infrared wavelength also extends[102,103] the cut-off energy in high harmonic generation and therefore intruiging prospects for ultrafast phenomena and generation of attosecond pulses as well as extension of precision spectroscopy in the extreme ultraviolet regions may be envisioned. Moreover, as sources of phase-stabilized femtosecond pulses, frequency combs have given a key to the production of attosecond pulses and the complete recovery of the electric field transients. Intense few-cycle long-wavelength laser systems[104] with carrier-envelope phase stabilization may thus benefit high field physics applications.

Many other applications, like dual-comb static optical coherence tomography[105] of tissues with low water content or dual-comb calibration[106] of rapidly swept continuous-wave mid-infrared lasers, might also emerge in the near future.

## 4. Conclusion

One decade after the first demonstration of near-infrared and visible frequency combs and the revolution they produced in optical frequency metrology and optical frequency synthesis, the rapidly evolving developments of frequency comb generator technology in the mid-infrared spectral region bring much confidence in the fact that mature and turn-key sources will become available in the near future and will stimulate much interest in novel applications. Fundamental and technological progress along several directions however still has to be achieved. Direct laser emission of few-cycle phase coherent pulses across the mid-infrared range, improved materials for efficient non-linear frequency conversion over a broad spectral bandwidth beyond 6 µm and octave-spanning low-phase noise microresonator-based frequency combs centered deeper in the mid-infrared may be perceived as being amongst the challenges to take up. The already explored applications, mostly benefiting spectroscopy of molecules, open intriguing prospects in frequency metrology and trace gas detection. They will readily take advantage of the progress in frequency comb technology. Most interestingly, one may hope for unforeseen groundbreaking measurement techniques and unexpected scientific discoveries, as the performance frontiers of these photonic tools is continuously advanced.

**Acknowledgments:** T.W.H. and N.P. acknowledge support by the European Associated Laboratory "European Laboratory for Frequency Comb Spectroscopy" and the Max Planck Foundation. A. S. acknowledges support from a Marie Curie IAPP program and the Swiss National Science Foundation. A. Amy-Klein, E.

Baumann, B. Darquié, P. de Natale, A. Foltynowicz-Matyba, F. Keilmann, T. J. Kippenberg, D. Mazzotti, N.R Newbury, K. Vodopyanov, C. Y. Wang and J. Ye are gratefully acknowledged for providing comments, data and figures.

**Additional information:** Correspondence and requests for materials should be addressed to N.P. (nathalie.picque@mpq.mpg.de)

## References

1   Udem, T., Holzwarth, R. & Hänsch, T.W. Optical frequency metrology. *Nature* **416**, 233-237 (2002).
2   Cundiff, S. T. & Ye, J. Colloquium: Femtosecond optical frequency combs. *Rev Mod Phys* **75**, 325-342 (2003).
3   Ye, J. & Cundiff, S. T.    (Springer, Berlin, Heidelberg, 2005).
4   Hall, J. L. Nobel Lecture: Defining and measuring optical frequencies. *Rev Mod Phys* **78**, 1279-1295, doi:Doi 10.1103/Revmodphys.78.1279 (2006).
5   Hänsch, T.W. Nobel Lecture: Passion for precision. *Rev Mod Phys* **78**, 1297-1309 (2006).
6   Diddams, S. A. The evolving optical frequency comb. *J Opt Soc Am B* **27**, B51-B62 (2010).
7   Page, R. H. *et al.* Cr2+-doped zinc chalcogenides as efficient, widely tunable mid-infrared lasers. *Ieee J Quantum Elect* **33**, 609-619 (1997).
8   Sorokina, I. T. Crystalline mid-infrared lasers. *Top Appl Phys* **89**, 255-349 (2003).
9   Sorokin, E., Naumov, S. & Sorokina, I. T. Ultrabroadband infrared solid-state lasers. *Ieee J Sel Top Quant* **11**, 690-712 (2005).
10  Mirov, S. B. *et al.* Progress in mid-IR Cr(2+) and Fe(2+) doped II-VI materials and lasers. *Opt Mater Express* **1**, 898-910 (2011).
11  Sorokin, E., Sorokina, I. T., Mandon, J., Guelachvili, G. & Picqué, N. Sensitive multiplex spectroscopy in the molecular fingerprint 2.4 mu m region with a Cr2+:ZnSe femtosecond laser. *Opt Express* **15**, 16540-16545 (2007).
12  Cizmeciyan, M. N., Cankaya, H., Kurt, A. & Sennaroglu, A. Kerr-lens mode-locked femtosecond Cr(2+):ZnSe laser at 2420 nm. *Opt Lett* **34**, 3056-3058 (2009).
13  Slobodtchikov, E. & Moulton, P. in *Lasers, Sources, and Related Photonic Devices, OSA Technical Digest (CD) (Optical Society of America, 2012)*    (2012).
14  Sorokin, E., Tolstik, N. & Sorokina, I. in *Lasers, Sources, and Related Photonic Devices, OSA Technical Digest (CD) (Optical Society of America, 2012)*    (2012).
15  Bernhardt, B. *et al.* Mid-infrared dual-comb spectroscopy with 2.4 mu m Cr(2+):ZnSe femtosecond lasers. *Appl Phys B-Lasers O* **100**, 3-8 (2010).
16  Fedorov, V. V. *et al.* 3.77-5.05-mu m tunable solid-state lasers based on Fe2+-doped ZnSe crystals operating at low and room temperatures. *Ieee J Quantum Elect* **42**, 907-917 (2006).
17  Frolov, M. P. *et al.* Laser Radiation Tunable within the Range of 4.35-5.45 Mu M in a Znte Crystal Doped with Fe(2+) Ions. *J Russ Laser Res* **32**, 528-536 (2011).
18  Kozlovsky, V. I. *et al.* Pulsed Fe(2+): ZnS laser continuously tunable in the wavelength range of 3.49-4.65 mu m. *Quantum Electron+* **41**, 1-3, doi (2011).
19  Pollnau, M. & Jackson, S. D. in *Top Appl Phys* Vol. 89 *Topics in Applied Physics* (eds I. Sorokina & K. Vodopyanov)  219-255 (Springer, 2003).
20  Nelson, L. E., Ippen, E. P. & Haus, H. A. Broadly Tunable Sub-500 Fs Pulses from an Additive-Pulse Mode-Locked Thulium-Doped Fiber Ring Laser. *Appl Phys Lett* **67**, 19-21 (1995).
21  Solodyankin, M. A. *et al.* Mode-locked 1.93 mu m thulium fiber laser with a carbon nanotube absorber. *Opt Lett* **33**, 1336-1338 (2008).
22  Kieu, K. & Wise, F. W. Soliton Thulium-Doped Fiber Laser With Carbon Nanotube Saturable Absorber. *Ieee Photonic Tech L* **21**, 128-130, doi:Doi 10.1109/Lpt.2008.2008727 (2009).
23  Haxsen, F., Wandt, D., Morgner, U., Neumann, J. & Kracht, D. Pulse characteristics of a passively mode-locked thulium fiber laser with positive and negative cavity dispersion. *Opt Express* **18**, 18981-18988 (2010).
24  Wang, Q., Geng, J. H., Jiang, Z., Luo, T. & Jiang, S. B. Mode-Locked Tm-Ho-Codoped Fiber Laser at 2.06 mu m. *Ieee Photonic Tech L* **23**, 682-684 (2011).

25    Phillips, C. R. *et al.* Supercontinuum generation in quasi-phase-matched LiNbO(3) waveguide pumped by a Tm-doped fiber laser system. *Opt Lett* **36**, 3912-3914 (2011).
26    Adler, F. & Diddams, S. A. High-power, hybrid Er:fiber/Tm:fiber frequency comb source in the 2 µm wavelength region. *Opt Lett* **37**, 1400-1402 (2012).
27    Coluccelli, N. *et al.* 1.6-W self-referenced frequency comb at 2.06 mu m using a Ho:YLF multipass amplifier. *Opt Lett* **36**, 2299-2301 (2011).
28    Hofstetter, D. & Faist, J. High performance quantum cascade lasers and their applications. *Top Appl Phys* **89**, 61-96 (2003).
29    Paiella, R. *et al.* Self-mode-locking of quantum cascade lasers with giant ultrafast optical nonlinearities. *Science* **290**, 1739-1742 (2000).
30    Wang, C. Y. *et al.* Mode-locked pulses from mid-infrared Quantum Cascade Lasers. *Opt Express* **17**, 12929-12943 (2009).
31    Fischer, C. & Sigrist, M. W. Mid-IR difference frequency generation. *Top Appl Phys* **89**, 97-140 (2003).
32    Maddaloni, P., Malara, P., Gagliardi, G. & De Natale, P. Mid-infrared fibre-based optical comb. *New J Phys* **8**, doi:Artn 262 Pii S1367-2630(06)29928-2 (2006).
33    Baumann, E. *et al.* Spectroscopy of the methane nu(3) band with an accurate midinfrared coherent dual-comb spectrometer. *Phys Rev A* **84**, doi:Artn 062513 (2011).
34    Foreman, S. M., Jones, D. J. & Ye, J. Flexible and rapidly configurable femtosecond pulse generation in the mid-IR. *Opt. Lett.* **28**, 370-372 (2003).
35    Foreman, S. M. *et al.* Demonstration of a HeNe/CH4-based optical molecular clock. *Opt Lett* **30**, 570-572 (2005).
36    Erny, C. *et al.* Mid-infrared difference-frequency generation of ultrashort pulses tunable between 3.2 and 4.8 mu m from a compact fiber source. *Opt Lett* **32**, 1138-1140 (2007).
37    Gubin, M. A. *et al.* Femtosecond fiber laser based methane optical clock. *Appl Phys B-Lasers O* **95**, 661-666 (2009).
38    Sell, A., Scheu, R., Leitenstorfer, A. & Huber, R. Field-resolved detection of phase-locked infrared transients from a compact Er:fiber system tunable between 55 and 107 THz. *Appl Phys Lett* **93**, 251107 (2008).
39    Keilmann, F., Gohle, C. & Holzwarth, R. Time-domain mid-infrared frequency-comb spectrometer. *Opt Lett* **29**, 1542-1544 (2004).
40    Schliesser, A., Brehm, M., Keilmann, F. & van der Weide, D. W. Frequency-comb infrared spectrometer for rapid, remote chemical sensing. *Opt Express* **13**, 9029-9038 (2005).
41    Gambetta, A., Ramponi, R. & Marangoni, M. Mid-infrared optical combs from a compact amplified Er-doped fiber oscillator. *Opt Lett* **33**, 2671-2673 (2008).
42    Keilmann, F. & Amarie, S. Mid-infrared frequency comb spanning an octave based on an Er fiber laser and difference-frequency generation. Journal of Infrared, Millimeter and Terahertz Waves **33**, 479-484 (2012).
43    Ruehl, A. *et al.* Widely-tunable mid-IR frequency comb source based on difference frequency generation. *arXiv:1203.2441, to be published in Optics Letters, DOC ID 164602* (2012).
44    Ebrahimzadeh, M. in *Top Appl Phys* Vol. 89 *Topics in Applied Physics* (eds I. Sorokina & K. Vodopyanov) 179-218 (Springer, 2003).
45    Vodopyanov, K. in *Top Appl Phys* Vol. 89 *Topics in Applied Physics* (eds I. Sorokina & K. Vodopyanov) 141-178 (Springer, 2003).
46    Reid, D. T., Gale, B. J. S. & Sun, J. Frequency comb generation and carrier-envelope phase control in femtosecond optical parametric oscillators. *Laser Phys* **18**, 87-103 (2008).
47    Sun, J. H., Gale, B. J. S. & Reid, D. T. Composite frequency comb spanning 0.4-2.4 mu m from a phase-controlled femtosecond Ti : sapphire laser and synchronously pumped optical parametric oscillator. *Opt Lett* **32**, 1414-1416 (2007).
48    Adler, F. *et al.* Phase-stabilized, 1.5 W frequency comb at 2.8–4.8 µm. *Opt Lett* **34**, 1330-1332 (2009).
49    Wong, S. T., Vodopyanov, K. L. & Byer, R. L. Self-phase-locked divide-by-2 optical parametric oscillator as a broadband frequency comb source. *J Opt Soc Am B* **27**, 876-882 (2010).
50    Leindecker, N., Marandi, A., Byer, R. L. & Vodopyanov, K. L. Broadband degenerate OPO for mid-infrared frequency comb generation. *Opt Express* **19**, 6304-6310 (2011).

51    Leindecker, N. *et al.* Octave-spanning ultrafast OPO with 2.6-6.1μm instantaneous bandwidth pumped by femtosecond Tm-fiber laser. *Opt. Express* **20**, 7046-7053 (2012).
52    Vodopyanov, K. L., Sorokin, E., Sorokina, I. T. & Schunemann, P. G. Mid-IR frequency comb source spanning 4.4-5.4 mu m based on subharmonic GaAs optical parametric oscillator. *Opt Lett* **36**, 2275-2277 (2011).
53    Savchenkov, A. A. *et al.* Low threshold optical oscillations in a whispering gallery mode CaF2 resonator. *Phys Rev Lett* **93**, doi:Artn 243905 (2004).
54    Del'Haye, P. *et al.* Optical frequency comb generation from a monolithic microresonator. *Nature* **450**, 1214-1217 (2007).
55    Kippenberg, T. J., Spillane, S. M. & Vahala, K. J. Kerr-nonlinearity optical parametric oscillation in an ultrahigh-Q toroid microcavity. *Phys Rev Lett* **93**, doi:Artn 083904 (2004).
56    Del'Haye, P. *et al.* Octave Spanning Tunable Frequency Comb from a Microresonator. *Phys Rev Lett* **107**, doi:Artn 063901 (2011).
57    Okawachi, Y. *et al.* Octave-spanning frequency comb generation in a silicon nitride chip. *Opt Lett* **36**, 3398-3400 (2011).
58    Domachuk, P. *et al.* Over 4000 nm bandwidth of mid-IR supercontinuum generation in sub-centimeter segments of highly nonlinear tellurite PCFs. *Opt Express* **16**, 7161-7168 (2008).
59    Yeom, D. I. *et al.* Low-threshold supercontinuum generation in highly nonlinear chalcogenide nanowires. *Opt Lett* **33**, 660-662 (2008).
60    Kuyken, B. *et al.* Mid-infrared to telecom-band supercontinuum generation in highly nonlinear silicon-on-insulator wire waveguides. *Opt Express* **19**, 20172-20181 (2011).
61    Qin, G. S. *et al.* Wideband supercontinuum generation in tapered tellurite microstructured fibers. *Laser Phys* **21**, 1115-1121 (2011).
62    Herr, T. *et al.* Universal Dynamics of Kerr Frequency Comb Formation in Microresonators. *arXiv: 1111.3071* (2011).
63    Savchenkov, A. A. *et al.* Tunable optical frequency comb with a crystalline whispering gallery mode resonator. *Phys Rev Lett* **101**, doi:Artn 093902 (2008).
64    Wang, C. Y. *et al.* Mid-Infrared Frequency Combs Based on Microresonators. *arXiv:1109.2716* (2011).
65    Daussy, C. *et al.* Long-Distance Frequency Dissemination with a Resolution of $10^{-17}$. *Phys Rev Lett* **94**, 203904 (2005).
66    Shelkovnikov, A., Butcher, R. J., Chardonnet, C. & Amy-Klein, A. Stability of the proton-to-electron mass ratio. *Phys Rev Lett* **100**, doi:Artn 150801 (2008).
67    Lemarchand, C. *et al.* Progress towards an accurate determination of the Boltzmann constant by Doppler spectroscopy. *New J Phys* **13**, doi:Artn 073028 (2011).
68    Darquié, B. *et al.* Progress Toward the First Observation of Parity Violation in Chiral Molecules by High-Resolution Laser Spectroscopy. *Chirality* **22**, 870-884, doi:Doi 10.1002/Chir.20911 (2010).
69    Malara, P., Maddaloni, P., Gagliardi, G. & De Natale, P. Absolute frequency measurement of molecular transitions by a direct link to a comb generated around 3-μm. *Opt Express* **16**, 8242-8249 (2008).
70    Amy-Klein, A. *et al.* Absolute frequency measurement of a SF6 two-photon line by use of a femtosecond optical comb and sum-frequency generation. *Opt Lett* **30**, 3320-3322 (2005).
71    Bielsa, F. *et al.* HCOOH high-resolution spectroscopy in the 9.18 μm region. *Journal of Molecular Spectroscopy* **247**, 41-46 (2008).
72    Gatti, D. *et al.* High-precision molecular interrogation by direct referencing of a quantum-cascade-laser to a near-infrared frequency comb. *Opt. Express* **19**, 17520-17527 (2011).
73    Bartalini, S. *et al.* Frequency-comb-referenced quantum-cascade laser at 4.4 μm. *Opt. Lett.* **32**, 988-990 (2007).
74    Giusfredi, G. *et al.* Saturated-Absorption Cavity Ring-Down Spectroscopy. *Phys Rev Lett* **104**, 110801 (2010).
75    Okubo, S., Nakayama, H., Iwakuni, K., Inaba, H. & Sasada, H. Absolute frequency list of the v(3)-band transitions of methane at a relative uncertainty level of 10(-11). *Opt Express* **19**, 23878-23888 (2011).
76    Vainio, M., Merimaa, M. & Halonen, L. Frequency-comb-referenced molecular spectroscopy in the mid-infrared region. *Opt. Lett.* **36**, 4122-4124 (2011).

77   Marian, A., Stowe, M. C., Lawall, J. R., Felinto, D. & Ye, J. United Time-Frequency Spectroscopy for Dynamics and Global Structure. *Science* **306**, 2063-2068, doi:10.1126/science.1105660 (2004).
78   Teets, R., Eckstein, J. & Hänsch, T.W. Coherent Two-Photon Excitation by Multiple Light Pulses. *Phys Rev Lett* **38**, 760-764 (1977).
79   Eckstein, J. N., Ferguson, A. I. & Hänsch, T.W. High-Resolution Two-Photon Spectroscopy with Picosecond Light Pulses. *Phys Rev Lett* **40**, 847-850 (1978).
80   Diddams, S. A., Hollberg, L. & Mbele. Molecular fingerprinting with the resolved modes of a femtosecond laser frequency comb. *Nature* **445**, 627-630 (2007).
81   Mandon, J., Guelachvili, G. & Picqué, N. Fourier transform spectroscopy with a laser frequency comb. *Nature Photonics* **4**, 55-57 (2009).
82   Bernhardt, B. *et al.* Cavity-enhanced dual-comb spectroscopy. *Nat Photonics* **4**, 55-57 (2010).
83   Thorpe, M. J. & Ye, J. Cavity-enhanced direct frequency comb spectroscopy. *Applied Physics B: Lasers and Optics* **91**, 397-414 (2008).
84   Adler, F., et al. . Cavity-Enhanced Direct Frequency Comb Spectroscopy: Technology and Applications. *Annual Review of Analytical Chemistry* **3**, 175-205 (2010).
85   Thorpe, M. J., et al. Broadband cavity ringdown spectroscopy for sensitive and rapid molecular detection. *Science* **311**, 1595-1599 (2006).
86   Griffiths, P. R. & De Haseth, J. A. *Fourier transform infrared spectroscopy*. 2nd edition edn, (John Wiley & Sons Inc., 2007).
87   Adler, F. *et al.* Mid-infrared Fourier transform spectroscopy with a broadband frequency comb. *Opt Express* **18**, 21861-21872 (2010).
88   Foltynowicz, A., Malowski, P., Fleisher, A. J., Bjork, B. & Ye, J. Cavity-enhanced optical frequency comb spectroscopy in the mid-infrared - application to trace detection of $H_2O_2$. *Appl Phys B-Lasers O* **in press**, Online First, 1 May 2012 (2012).
89   Amarie, S. & Keilmann, F. Broadband-infrared assessment of phonon resonance in scattering-type near-field microscopy. *Phys. Rev. B* **83**, 045404 (2011).
90   Ganz, T., Brehm, M., von Ribbeck, H.G., van der Weide, D.W.; Keilmann F. Vector frequency-comb Fourier-transform spectroscopy for characterizing metamaterials. *New J Phys* **10**, 123007 (2008).
91   Brehm M., Schliesser A. & Keilmann, F. Spectroscopic near-field microscopy using frequency combs in the mid-infrared. *Opt. Express* **14**, 11222-11233 (2006).
92   Coddington, I., Swann, W.C. & Newbury, N.R. . Coherent multiheterodyne spectroscopy using stabilized optical frequency combs. *Phys. Rev. Lett* **100**, 013902 (2007).
93   Zolot, A. M. *et al.* Direct-comb molecular spectroscopy with accurate, resolved comb teeth over 43 THz. *Optics Letters* **37**, 638-640 (2012).
94   Ideguchi, T., Poisson, A., Guelachvili, G., Picqué, N. & Hänsch, T.W. Adaptive real-time dual-comb spectroscopy. *arXiv :1201.4177* (2012).
95   Ideguchi, T., Bernhardt, B., Guelachvili, G., Hänsch, T.W. & Picqué, N. Raman-induced Kerr effect dual-frequency-comb spectroscopy. *arXiv:* (2012).
96   Zhang, Z. *et al.* Asynchronous midinfrared ultrafast optical parametric oscillator for dual-comb spectroscopy. *Opt. Lett.* **37**, 187-189 (2012).
97   Cundiff, S. T. & Weiner, A. M. Optical arbitrary waveform generation. *Nat Photon* **4**, 760-766 (2010).
98   Steinmetz, T. *et al.* Laser frequency combs for astronomical observations. *Science* **23**, 1335 (2008).
99   Li, C.-H. *et al.* A laser frequency comb that enables radial velocity measurements with a precision of 1 cms−1. *Nature* **452**, 610-612 (2008).
100  Mayor, M. & Queloz, D. A Jupiter-mass companion to a solar-type star. *Nature* **378**, 355-359 (1995).
101  Figueira, P. *et al.* Radial velocities with CRIRES. Pushing precision down to 5-10 m/s. *Astronomy and Astrophysics* **51**, A55 (2010).
102  Sheehy, B. *et al.* High Harmonic Generation at Long Wavelengths. *Phys Rev Lett* **83**, 5270-5273 (1999).
103  Krause, J. L., Schafer, K. J. & Kulander, K. C. High-order harmonic generation from atoms and ions in the high intensity regime. *Physical Review Letters* **68**, 3535-3538 (1992).

104     Silva, F., Bates, P. K., Esteban-Martin, A., Ebrahim-Zadeh, M. & Biegert, J. High-average-power, carrier-envelope phase-stable, few-cycle pulses at 2.1µm from a collinear BiB3O6 optical parametric amplifier. *Opt. Lett.* **37**, 933-935 (2012).
105     Lee, S. J., Widiyatmoko, B., Kourogi, M. & Ohtsu, M. Ultrahigh scanning speed optical coherence tomography using optical frequency comb generators. *Jpn J Appl Phys 2* **40**, L878-L880 (2001).
106     Giorgetta, F. R., Coddington, I., Baumann, E., Swann, W. C. & Newbury, N. R. Fast high-resolution spectroscopy of dynamic continuous-wave laser sources. *Nat Photonics* **4**, 853-857 (2010).
107     Rothman, L. S. *et al.* The HITRAN 2008 molecular spectroscopic database. *Journal of Quantitative Spectroscopy and Radiative Transfer* **110**, 533-572 (2009).